\documentclass[aps,prb,reprint,amsmath,amssymb,superscriptaddress,notitlepage,longbibliography]{revtex4-2}
\usepackage[utf8]{inputenc}
\usepackage{bm}
\usepackage{enumerate}
\usepackage{graphicx}
\usepackage{cancel}
\usepackage{xcolor}
\usepackage{verbatim}
\usepackage{hyperref}
\usepackage{amsmath}
\usepackage{mathtools}
\usepackage{amsfonts}
\usepackage{physics}

\hypersetup{colorlinks=true,
linkcolor=blue,
citecolor=blue,
urlcolor=blue,
filecolor=blue
}

\def\bra#1{\langle{#1}|}
\def\ket#1{|{#1}\rangle}

\begin{abstract}
Protecting superconducting qubits from low-frequency noise by operating them on dynamical sweet-spot manifolds has proven to be a promising setup, theoretically as well as experimentally. These dynamical sweet spots are induced by an externally applied Floquet drive, and various drive forms have been studied in different types of qubits. In this work we study the effects of using two-tone drives on the applied magnetic flux of the form $\phi_{ac}(t)=\phi_m\cos(m\omega_\mathrm{d} t)+\phi_n\cos(n\omega_\mathrm{d} t+\varphi)$, where $m,n \in \mathbb{N}_{>0}$, on the coherence times of fluxonium qubits. The optimal drive parameters are found through analysis using perturbation theory and numerical calculations. We show that this type of drive allows for more tunability of the quasi-energy spectrum, creating higher and wider peaks of the dephasing time without affecting the relaxation times too strongly. Further we show that the second commensurable drive tone can be used to implement an improved phase gate compared to implementations with a single tone, supported by Monte Carlo simulations.
\end{abstract}

\begin{document}

\title{Optimization of Floquet fluxonium qubits with commensurable two-tone drives}

\author{Joachim Lauwens}

\affiliation{Department of Physics and Astronomy, KU Leuven, Celestijnenlaan 200D, 3001 Leuven, Belgium.}

\author{Kristof Moors}
\affiliation{Imec, Kapeldreef 75, 3001 Heverlee, Belgium.}

\author{Bart Sor\'ee}
\affiliation{Department of Electrical Engineering, KU Leuven, Kasteelpark Arenberg 10, 3001 Leuven, Belgium.}
\affiliation{Department of Physics, University of Antwerp, Groenenborgerlaan 171, 2020 Antwerp, Belgium.}

\date{\today}


\maketitle

\section{Introduction}
\label{introduction}

Fluxonium qubits have been measured to have long coherence times compared to other superconducting qubits, with experiments measuring $T_2$ reaching as high as $1.48$ ms \cite{PhysRevLett.130.267001, PhysRevX.9.041041}. These high coherence times can be attributed to their insensitivity to charge noise \cite{doi:10.1126/science.1175552} and operating points at anticrossings, which protect the fluxonium from dephasing due to low-frequency noise in the applied magnetic flux \cite{PhysRevX.9.041041, PhysRevX.11.011010, PhysRevApplied.20.034016}. However, away from these anticrossing sweet spots the dephasing time quickly deteriorates \cite{e5d63f82c2a64f1ba376efd945b0ab65}, limiting the tunability of the qubit's energy spectrum. 

Applying a time-periodic drive to superconducting qubits has been proposed as a way of increasing their coherence times in the presence of 1/$f$ magnetic flux noise \cite{huang2021engineering,nguyen2024programmable,gandon2022engineering,thibodeau2024floquet,tsuji2023floquet}, by operating them on so-called dynamical sweet spots \cite{frees2019adiabatic, 2013Natur.494..211P, wang2024quantum}, which exhibit more tunability than their static counterparts. These sweet spots are located at anticrossings between the quasi-energy levels of Floquet states, and form manifolds in the space of drive parameters. This makes them especially interesting for implementing gate operations since the drive can be modulated while retaining sweet-spot operation.

As shown in Ref.~\cite{PhysRevApplied.14.054033}, fluxonium qubits have both relatively high dephasing times as well as relaxation times at dynamical sweet-spots. In fluxonium qubits, the drive is usually achieved by applying a time-periodic magnetic flux to the closed loop in the circuit formed by the Josephson junction and the inductor. Existing proposals use modulation of the magnetic flux of a pure frequency $\omega_\mathrm{d}$ in fluxonium and transmon qubits, two frequencies $\omega_\mathrm{d}$ and $n\omega_\mathrm{d}$, with $n \in \mathbb{N}$ in transmons \cite{PRXQuantum.3.020337}, or a pure frequency with a complex phase \cite{Cheng_2022}. 

In this paper we study a periodically driven fluxonium qubit with a drive consisting of two tones with a relative phase difference, $\phi_{ac}(t)=\phi_m\cos(m\omega_\mathrm{d} t)+\phi_n\cos(n\omega_\mathrm{d} t+\varphi)$, where $m,n \in \mathbb{N}_{>0}$ and $\gcd(m,n)=1$ such that the period of the Hamiltonian $T_d$ is $2\pi/\omega_\mathrm{d}$. Using this type of drive to define the Floquet states allows for more flexibility in finding optimal driving parameters with respect to the coherence times. In addition, the second drive tone can also be used to implement a phase gate in a Floquet qubit where the states are defined by a single drive tone. This results in more tunability in the quasi-energy splitting and an improved gate fidelity. 

Here is a brief overview of the paper. We start in Sec.~\ref{section 2} by giving an overview of the system under consideration and analyzing the anticrossings in the quasi-energy spectrum perturbatively. Next, we study the tunability of these anticrossings when varying the amplitudes and frequencies of the drive. In Sec.~\ref{section 3} the expressions for the dephasing and relaxation times for the Floquet states are derived, taking into account noise on the static magnetic flux and on the drive amplitudes. In Sec.~\ref{section 4} optimal drive parameters are derived from the perturbative expression for the gap size at an anticrossing. The optimal choices for $m,n$ and $\varphi$ are explained. Further, in Sec.~\ref{section 5}, a proposal for a two-tone phase gate is explored using numerical simulations. We conclude in Sec.~\ref{Conclusion}.

\section{Fluxonium with a two-tone drive}

\label{section 2}

\begin{figure*}
    \centering
    \includegraphics[]{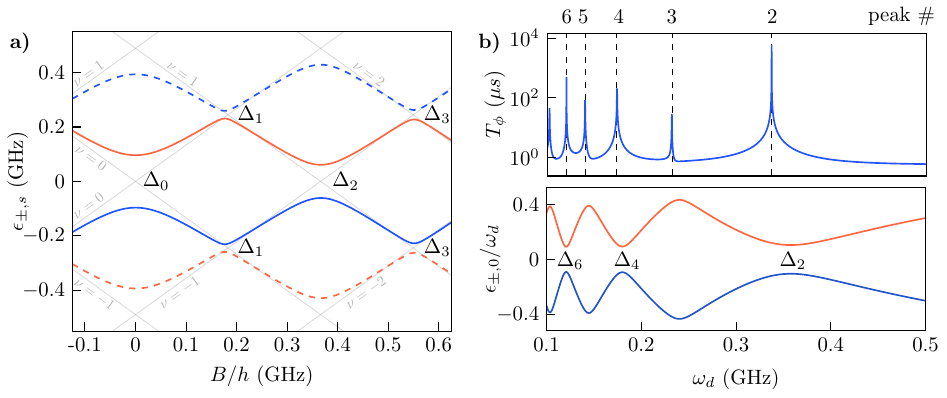}
    \caption{\textbf{a)} Quasi-energy spectrum for a range of applied static magnetic fluxes $\phi_{\mathrm{dc}}$. The blue (red) lines are the quasi-energies associated with the $\ket{v_{+,\nu}(t)}$ $(\ket{v_{-,\nu}(t)})$ Floquet modes. The solid lines show the quasi-energies closest to zero, while the dashed lines show the second-lowest Floquet copies. The $\Delta_s$ represent the magnitudes of the spectral gap at the respective anticrossings, where $s$ represents the `photon number' difference between the relevant states. Grey lines show the quasi-energies when $\Delta=0$, with the $s$-values of the corresponding Floquet modes annotated. \textbf{b)} Pure dephasing time for a range of drive frequencies, and the corresponding quasi-energy spectrum below. $T_\phi$-peaks are present in the vicinity of each anticrossing, with the peak numbers equal to $s$ in \textbf{a)}. The parameters used in this plot are $\Delta/h=0.28\text{ GHz}$ and $B/h=0.65\text{ GHz}$, which correspond to fluxonium parameters $E_\mathrm{C} = 0.7$ GHz, $E_\mathrm{L} = 1.3$ GHz, $E_\mathrm{J} = 4.0$ GHz (see App.~\ref{app:qubithamiltonian}). Further, $\omega_\mathrm{d} = 0.49$ GHz, $A_m/h = 0.42$ GHz, $A_n/h = 0.50$ GHz, $(m,n,\varphi)=(2,3,0)$.}
    \label{fig:energiesandpeaks}
\end{figure*}

The system under investigation is a fluxonium driven by an oscillating magnetic flux through the loop formed by its inductor and Josephson junction. We consider the Hamiltonian in the two-level approximation,
\begin{equation}
\begin{aligned}
    \hat{H}(t) = \frac{\Delta}{2}\hat{\sigma}_x + \bigg(\frac{B}{2} &+ A_m\cos(m\omega_\mathrm{d} t) \\
    &+ A_n\cos(n\omega_\mathrm{d} t + \varphi)\bigg)\hat{\sigma}_z,
    \label{eq:system}
\end{aligned}
\end{equation}
 with $\hat{\sigma}_z$ and $\hat{\sigma}_x$ Pauli matrices. The parameters $\Delta$ and $B$ are related to the circuit elements and static flux of the fluxonium. $A_m$ and $A_n$ are the drive amplitudes, related to the flux drive introduced in Sec.~\ref{introduction} (see App.~\ref{app:qubithamiltonian} for details). Since this Hamiltonian is time periodic with period $2\pi/\omega_\mathrm{d}$, its dynamics can be described by Floquet modes $\ket{v_{\pm, \nu}(t)}$ \cite{PhysRevLett.115.133601} and their corresponding quasi-energies $\epsilon_{\pm, \nu}$ \cite{PhysRev.138.B979, GRIFONI1998229, CHU20041, Silveri_2017, ASENS_1883_2_12__47_0}. These are found by solving the Floquet equation
\begin{equation}
    \bigg(\hat{H}(t) - \frac{\partial}{\partial t}\bigg)\ket{v_{\pm, \nu}(t)} = \epsilon_{\pm, \nu}\ket{v_{\pm, \nu}(t)},
\end{equation}
which has two independent solutions, indicated by $\pm$. The integer $s$ represents the `photon number' of the Floquet mode. The quasi-energies of the two solutions are thus not unique and satisfy $\epsilon_{\pm, \nu} = \epsilon_{\pm,0} + \nu\hbar\omega_\mathrm{d}$, resulting in an energy spectrum with an infinite number of copies of $\epsilon_{\pm, 0}$ offset by $\hbar\omega_\mathrm{d}$.

Fig.~\ref{fig:energiesandpeaks}(a) shows a section of the energy spectrum of the driven fluxonium for $(m, n, \varphi) = (2, 3, 0)$, in which several anticrossings between $\epsilon_{+, i}$ and $\epsilon_{-, j}$ are visible. At these locations the qubit is protected against dephasing due to the low-frequency noise on the static magnetic flux $\phi_{dc}$ as illustrated in Fig.~\ref{fig:energiesandpeaks}(b), which shows increased dephasing times $T_\phi$ in the regions of the anticrossings. The $T_\phi$-peaks are labeled by a peak number $s$ that corresponds to the `photon number' difference $|\nu-\nu'|$ between the relevant Floquet states at the corresponding anticrossing. It is thus useful to analyze the magnitudes of these anticrossings, and their dependence on the drive parameters.

In the strong-drive limit, $\frac{\Delta}{2}\hat{\sigma}_x$ can be seen as a perturbation \cite{huang2021engineering}, as detailed in App.~\ref{App:GapSize}, and an expression can be derived for the gap size at the anticrossings between quasi-energies $\epsilon_{+,0}$ and $\epsilon_{-,s}$. Using first-order perturbation theory, the quasi-energy splitting $\Delta\epsilon_s(B,\omega_\mathrm{d}) \equiv \epsilon_{+,0} - \epsilon_{-,s}$ can be approximated, in the vicinity of an anticrossing, by $\Delta\epsilon_s(B, \omega_\mathrm{d})\approx\sqrt{\Delta^2_s + (B - s\omega_\mathrm{d})^2}$ with $\Delta_s$ the gap size at the anticrossing ($B=s\omega_\mathrm{d}$),
\begin{equation}
    \Delta_s = \Delta\abs{\sum_{(k,k')}J_k\bigg(\frac{2A_m}{m\omega_\mathrm{d}}\bigg)J_{k'}\bigg(\frac{2A_n}{n\omega_\mathrm{d}}\bigg)e^{ik'\varphi}}.
    \label{eq:delta_s}
\end{equation}
The summation in this expression runs over the pairs $(k, k')$, which form integer solutions to the linear Diophantine equation 
\begin{equation}
    mk+nk'+s=0,
    \label{eq:diophantine}
\end{equation}
and $J_k(z)$ are Bessel functions of the first kind. Note that Eq.~\eqref{eq:delta_s} reduces to $\Delta_s = \Delta\abs{J_s\big(\frac{2A}{\omega_\mathrm{d}}\big)}$ for a drive with a single frequency ($m=1$, $A_m = A$, $A_n=0$), as derived in Ref.~\cite{huang2021engineering}.

The magnitude of the anticrossing shows a clear dependence on all of the driving parameters, and their influence on the coherence times is studied in the next sections. In the remainder of the text, the drive form will be specified using the list ($m$, $n$, $\varphi$, $s$), where $m$, $n$, and $\varphi$ are fixed drive parameters and $s$ is the peak number under consideration. Note that $s$ specifies the base drive frequency to be tuned at the relevant peak in the dephasing time, $\omega_\mathrm{d}=\omega_\mathrm{d}^\mathrm{peak}$.

\begin{figure*}
    \centering
    \includegraphics[width=\textwidth]{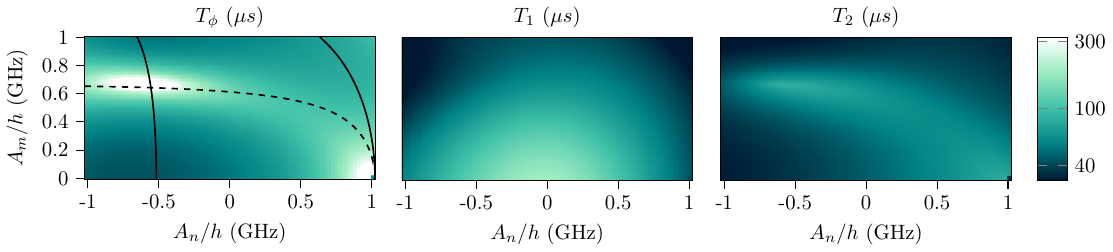}
    \caption{The coherence times as a function of two-tone drive amplitudes $A_m$ and $A_n$ when driving the fluxonium with parameters $(m,n,\varphi,s)=(1,2,0,1)$. From left to right the plots show the pure dephasing time, relaxation time and total dephasing time. Black lines in the left plot indicate the location of $A_m$ and $A_n$ sweet spots, where they cross are the dephasing triple sweet spots. In these plots $\omega_\mathrm{d}$ was tuned to $\omega_\mathrm{d}^\mathrm{peak}$ at each point ($A_m$, $A_n$) to remain at the $T_2$-peak, with respect to $\omega_\mathrm{d}$.}
    \label{fig:coherence times}
\end{figure*}

\section{Decoherence rates}
\label{section 3}

While fluxoniums are insensitive to charge noise, they are still affected by various other sources of noise \cite{PhysRevLett.97.167001, PhysRevB.93.104518, PhysRevX.11.011010}. Here we consider $1/f$-noise in the applied static magnetic flux as well as the drive amplitudes, and dielectric loss. The effects of these noise sources are modeled using baths coupled to the qubit through an interaction Hamiltonian \cite{breuer2002theory, huang2021engineering, PhysRevApplied.14.054033},
\begin{equation}
\begin{aligned}
    H_\text{int} = \hat{\sigma}_z(\hat{\beta}^\mathrm{dc}+\hat{\beta}^\mathrm{dl})&+\cos(m\omega_\mathrm{d}t)\hat{\sigma}_z\hat{\beta}^\mathrm{ac, m} \\&+ \cos(n\omega_\mathrm{d}t+\varphi)\hat{\sigma}_z\hat{\beta}^\mathrm{ac, n},
\end{aligned}
\label{Hint}
\end{equation}
with $\hat{\beta}^\mathrm{dc}$ and $\hat{\beta}^\mathrm{dl}$ the bath operators corresponding to static flux noise and dielectric loss \cite{PhysRevLett.120.150503}, respectively, and $\hat{\beta}^\mathrm{ac,m}$ and $\hat{\beta}^\mathrm{ac,n}$ the bath operators corresponding to noise on the drive amplitudes. The noise spectra describing these noise sources are detailed in App.~\ref{app:noise}.

The coupling of the baths to the qubit results in decoherence of the qubit state, quantified by the dephasing and relaxation rates. These rates can be expressed as a sum of decoherence rates for each noise source. The relaxation rate is the sum of the absorption ($\gamma_+$) and emission ($\gamma_-$) rates, $\gamma_1 = \gamma_+ + \gamma_-$, with
\begin{align}
    \gamma_{\pm} = \Gamma_\pm^\text{dc}+\Gamma_\pm^\text{dl}+\Gamma_\pm^\text{ac,m}+\Gamma_\pm^\text{ac,n}.
\end{align}
The corresponding relaxation time is then $T_1 = 1/\gamma_1$. The pure dephasing rate is similarly
\begin{align}
    \gamma_{\phi} = \Gamma_\phi^\text{dc}+\Gamma_\phi^\text{dl}+\Gamma_\phi^\text{ac,m}+\Gamma_\phi^\text{ac,n},
\end{align}
with pure dephasing time $T_\phi = 1/\gamma_\phi$, and the total dephasing time is given by $1/T_2 = 1/(2T_1)+1/T_\phi$. The full expressions for these rates are given in App.~\ref{dec_rates}, and depend strongly on the noise filter weights $g_{k}^\pm$ and $g_{k}^\phi$. These weights are the Fourier transforms of the coupling matrix elements \cite{huang2021engineering},
\begin{equation}
    g_{k}^\pm = \frac{\omega_\mathrm{d}}{2\pi}\int_0^{\frac{2\pi}{\omega_\mathrm{d}}}\dd{t}\,e^{i k\omega_\mathrm{d} t}\bra{v_{\pm,0}(t)}\hat{\sigma}_z\ket{v_{\mp,0}(t)},
    \label{gterms1}
\end{equation}
\begin{equation}
\begin{aligned}
    g_{k}^\phi = \frac{\omega_\mathrm{d}}{4\pi}\int_0^{\frac{2\pi}{\omega_\mathrm{d}}}\dd{t}\,e^{i k\omega_\mathrm{d} t}\bigg(&\bra{v_{+,0}(t)}\hat{\sigma}_z\ket{v_{+,0}(t)}\\-&\bra{v_{-,0}(t)}\hat{\sigma}_z\ket{v_{-,0}(t)}\bigg),
    \label{gterms2}
\end{aligned}
\end{equation}
with $k\in\mathbb{Z}$.

The system exhibits three kinds of dephasing sweet-spots, one for each type of $1/f$-noise present. As an example, the pure dephasing rate due to noise on $A_m$ can be expressed as 
\begin{equation}
    \Gamma_\phi^\text{ac,m} = \sum_{k\in\mathbb{Z}}\frac{1}{2}\abs{g_{k-m}^\phi+g_{k+m}^\phi}^2S_{m}^\text{ac}(k\omega_\mathrm{d})
\end{equation}
Due to the strong peak of low-frequency $1/f$-noise around $\omega=0$, it is clear that the filter weights with $k=0$ have the strongest influence on the pure dephasing time. Sweet spots with respect to noise on $A_m$ thus correspond to points in the drive parameter space at which $g_{-m}^\phi = g_{m}^\phi=0$ \cite{PhysRevApplied.12.054015}. Away from this sweet spot, $\Gamma^\mathrm{ac,m}_{\phi}$ experiences a strong contribution from the peak in the low-frequency drive noise at $\omega=0$. Similar reasoning holds for sweet spots with respect to noise on $B$ ($g_0^\phi=0$) and $A_n$ ($g_{-n}^\phi = g_n^\phi=0$). 
Further, in static qubits, dephasing arises from noise-induced fluctuations in the energy splitting between eigenstates. Similarly, in a driven fluxonium, dephasing arises from noise that modulates the quasi-energy splitting. The filter weights are related to this quasi-energy splitting, since it can be shown that a small perturbation $B \rightarrow B + \delta B$ induces a change of the quasi-energy splitting $\frac{\partial\Delta\epsilon}{\partial B} = g_{0, \phi}$, up to first order in $\delta B$ \cite{huang2021engineering}. While this explains why the dephasing sweet-spots are located at anticrossings in the quasi-energy spectrum, the exact height of the $T_\phi$-peaks is limited by the remaining filter weights.

\begin{figure*}[t]  
    \centering
    \includegraphics[width=\textwidth]{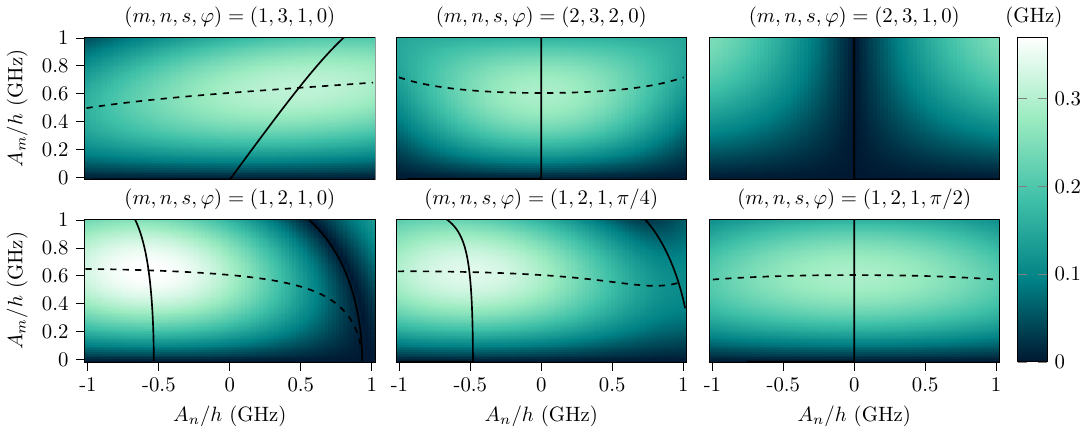}
    \caption{The anticrossing gap size $\Delta_s$ as a function of two-tone drive amplitudes $A_m$ and $A_n$ for different sets of drive parameters ($m,n,s,\varphi$). In the upper left plot, $(m,n,\varphi, s) = (1, 3, 0, 1)$ are considered, and the local maximum is located around $A_n/h=0.5\text{ GHz}$. In the upper middle plot, the drive parameters are $(2, 3, 0, 2)$. Here, the local maximum is located at $A_n/h=0\text{ GHz}$, since $s\pm n=2\pm3$ is not evenly divisible by $m=2$ as explained in App.~\ref{mnchoice}. In the upper right plot, the drive parameters are $(2,3, 0, 1)$. When increasing $A_n$ starting from zero, $\Delta_s$ increases. However, as detailed in App.~\ref{mnchoice}, there is no anticrossing for these drive parameters when $s=1$. The lower plots show $\Delta_s$ with $(m,n, \varphi, s)=(1,2, \varphi, 1)$ and different $\varphi$. When $\varphi=0$, there is a clear peak near $A_n/h=-0.5\text{ GHz}$, and $\partial\Delta_s/\partial A_n\neq0$ as explained in the text. Increasing the phase difference $\varphi$ causes $\partial\Delta_s/\partial A_n$ to decrease until it is zero at $\varphi=\pi/2$, since $\cos(\pi/2)=0$. At this point the local maximum has been shifted to $A_n=0$, and turning on the second drive frequency will lower $\Delta_s$. The dashed and solid black lines indicate the locations of $A_m$ and $A_n$ sweet spots, respectively. }
    \label{fig:wide_figure}
\end{figure*}

The use of two drive frequencies yields more control over these $g_{i,\phi}$-terms. In App.~\ref{noisy_drive}, the following relation is derived, 
\begin{equation}
    \frac{\partial\Delta\epsilon}{\partial A_i} = g_{i, \phi}+g_{-i, \phi},
    \label{eq:dEdA}
\end{equation}
which relates the change in quasi-energy splitting with drive amplitudes to the dephasing filter weights. Note that this relation was already derived for $i=1$ in \cite{PhysRevApplied.14.054033}. Thus, by using two drive tones, dephasing triple sweet spots are attainable, since these correspond to a local extremum in $\Delta\epsilon$ with respect to $B$, $A_m$ and $A_n$. The positive effect of triple sweet spots on the pure dephasing time is twofold. First, at those operating points the strong peak in the low-frequency noise spectra loses influence. Second, since five of the filter weights are zero, the height of the $T_\phi$-peak is increased further.

An important limitation of sweet-spot engineering is the redistribution of the filter weights. These weights satisfy the following conservation relation \cite{huang2021engineering, PhysRevB.77.174509}, 
\begin{equation}
    \sum_{k\in\mathbb{Z}}|g_k^+|^2+|g_k^-|^2+2|g_k^\phi|^2 = 2,
\end{equation}
and, as a result, lowering the dephasing filter weights causes an increase in the relaxation filter weights. As an example, the coherence times at the first peak of a driven fluxonium with drive parameters $(m,n,s,\varphi)=(1,2,1,0)$ are shown in Fig.~\ref{fig:coherence times}. The black lines cross at the dephasing triple sweet spot, at which the dephasing time indeed exhibits a maximum. Depending on the drive strengths, the coherence time of the qubit state is limited by either $T_1$ or $T_2$.

\section{Influence of drive parameters on the dephasing time}
\label{section 4}

With the aim of using the second drive to maximize the dephasing time, it is useful to study the location of the dephasing triple sweet spot for different drive parameters ($m, n, s, \varphi$). Since, at a triple sweet spot $g^\phi_{0}=g^\phi_{\pm m}=g^\phi_{\pm n}=0$, the pure dephasing time is expected to be maximal in such a region. Assuming the base frequency of the drive to stay tuned at an anticrossing ($\omega_\mathrm{d} =\omega_\mathrm{d}^\mathrm{peak}$ such that $g^\phi_0=0$) and a constant static flux, the quasi-energy splitting equals the gap size at the anticrossing. In that case, Eq.~\eqref{eq:dEdA} can be rewritten as $\partial \Delta_s/\partial A_i = g_{i, \phi}+g_{-i,\phi}$. Dephasing triple sweet-spots thus coincide with local extrema of $\Delta_s$ with respect to $A_m$ and $A_n$. The first-order correction of the second drive tone is beneficial if ${\partial \Delta_s}/{\partial A_n} \neq 0$ at $A_n=0$, which we refer to as the \emph{derivative condition}. When this condition is met, $\Delta_s$ can be tuned linearly with the amplitude of the second drive tone, and thus $T_\phi^\mathrm{peak}$ can be increased up until a local maximum ($A_n$-sweet spot) is reached.

In this section, we look for combinations of $(m,n,\varphi,s)$ such that the derivative condition is met. To this end, the derivative condition is evaluated using Eq.~\eqref{eq:delta_s},
\begin{equation} \label{eq:deriv}
\begin{split}
    &\frac{\partial\Delta_s}{\partial A_n}\Bigg|_{A_n=0} \\
    &\quad = \Delta\bigg[ J_{k_+}\bigg(\frac{2A_m}{m\omega_\mathrm{d}^\mathrm{peak}}\bigg) - J_{k_{-}}\bigg(\frac{2A_m}{m\omega_\mathrm{d}^\mathrm{peak}}\bigg) \bigg]\cos(\varphi),
\end{split}
\end{equation}

\begin{figure*}[htb]
    \centering
    \includegraphics[width=\textwidth]{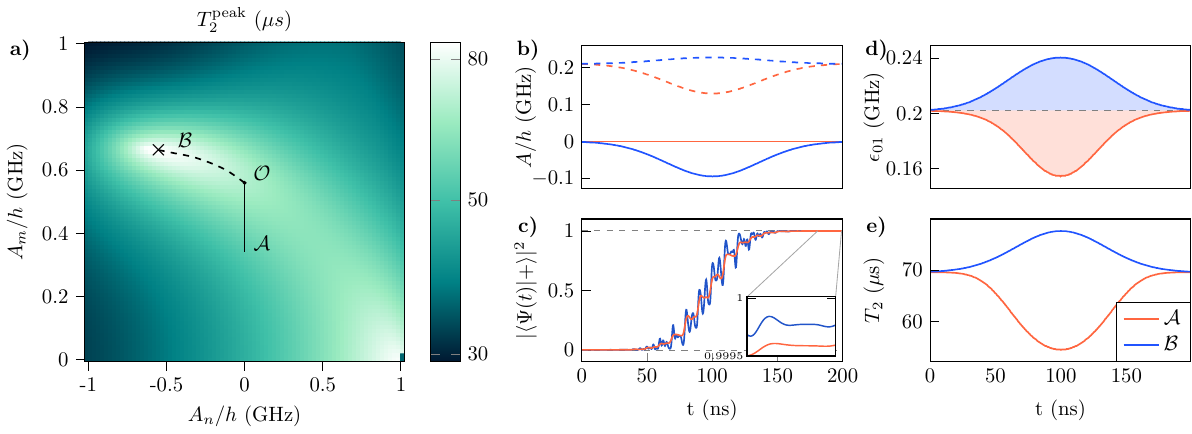}
    \caption{\textbf{a)} Maximal value of $T_2$  for a range of drive amplitudes, with $(m,n,s) = (1,2,1)$. There is a clear peak in the vicinity of the triple sweet-spot, indicated by the cross, which can be used to perform a phase gate. \textbf{b)} Modulation of the drive amplitudes during the phase gate pulse. The dashed lines represent $A_m$, the solid lines $A_n$. Modulating along path $\mathcal{A}$ (orange) keeps $A_n$ zero throughout the pulse. The modulation amplitude of $A_m$ when using path $\mathcal{A}$ is equal to the modulation amplitude of $A_n$ when using path $\mathcal{B}$ (blue). \textbf{c)} Overlap of the gated qubit state $\ket{\Psi(t)}$ and the state $\ket{+}$ during the phase-gate pulse. \textbf{d)} Change in the quasi-energy splitting during the phase-gate pulse. The gate parameters were chosen such that both gates result in a $\pi$-rotation of the initial state. \textbf{e)} Variation in $T_2$ during the gate pulse. Using path $\mathcal{A}$ results in a decrease of $T_2$, while using path $\mathcal{B}$ results in an increase.}
    \label{fig:PhaseGate}
\end{figure*}

with $k_+$ and $k_-$ the solutions of Eq.~\eqref{eq:diophantine} for $k$ when $k'=1$ and $k'=-1$, respectively. We assume $\omega_\mathrm{d}^\mathrm{peak}$ to be constant, which is a good approximation in the strong-drive regime \cite{huang2021engineering, PhysRevA.75.063414}. If no solutions for $k_+$ and $k_-$ exist, Eq.~\eqref{eq:deriv} equals zero and the quasi-energy splitting exhibits a local extremum at $A_n=0$. As explained in App.~\ref{mnchoice}, $k_\pm = (s\mp n)/m$, thus integer solutions for $k_\pm$ exist only when $m$ evenly divides $s\mp n$.

There are two distinct situations in terms of $m,n$ and $s$ which affect Eq.~\eqref{eq:deriv}.
First, if $m=1$ and $n\neq1$, the derivative condition is met for every peak number $s$. Examples of $\Delta_s$ using these drive parameters are shown in the upper and lower left plots in Fig.~\ref{fig:wide_figure}, for drive parameters $(1,3,0,1)$ and $(1,2,0,1)$ respectively. In these plots the triple sweet spot, indicated by the crossing point of the solid and dashed lines, is clearly located away from $A_n=0$, and using the second drive tone results in an increase of the peak dephasing times.
Second, if both $m\neq1$ and $n\neq1$, the derivative condition is not met for every peak number. More specifically, the condition is not met for the peak numbers that are integer multiples of $m$ (see App.~\ref{mnchoice}), which is illustrated in the upper middle plot in Fig.~\ref{fig:wide_figure} for the drive parameters $(2,3,0,2)$. When the condition is met, so when $s$ is not an integer multiple of $m$, there is no peak in the dephasing time at $A_n=0$. This is explained in App.~\ref{mnchoice}, and an example is shown in the upper right plot in Fig.~\ref{fig:wide_figure} for drive parameters $(2,3,0,1)$.
Finally, the effect of the relative phase shift $\varphi$ is clearly visible from Eq.~\eqref{eq:deriv}. When there is no phase shift the derivative is maximal, and increasing $\varphi$ causes the the derivative to decrease to zero when $\varphi=\pi/2$, as illustrated in the three bottom plots in Fig.~\ref{fig:wide_figure}.
In conclusion, the drive parameters that are of interest with increasing $T_\phi^\mathrm{peak}$ in mind are $m=1$, $n\neq1$ and $\varphi=0$. For these parameters $T_\phi^\mathrm{peak}$ increases for every peak number $s$.

\section{Two-tone phase gate}

\label{section 5}


A second driving tone can also be used to implement single-qubit gates. In Ref.~\cite{huang2021engineering}, a second drive tone resonant with the quasi-energy difference between the Floquet modes is used to induce Rabi oscillations, effectively implementing an X-gate. Alternatively, a phase gate can be implemented by modulating the drive parameters along a sweet-spot manifold, since this modulates the quasi-energy splitting. The second drive tone allows for more tunability while remaining on the sweet spot manifold, which can be used to improve the fidelity of the phase gate. Fig.~\ref{fig:PhaseGate}(a) shows the peak value of $T_2$ as a function of $A_m$ and $A_n$, with $(m,n,\varphi,s)=(1,2,0,1)$ and $\omega_\mathrm{d}$ tuned at the peak value. As expected from the analysis of $T_\phi$ in Sec.~\ref{section 4}, for this drive parameter set the total dephasing time can be increased by using the second drive tone. The maximal value of $T_2$ is obtained in the region near the dephasing triple sweet-spot. 


We consider two paths in the driving parameter space, shown in the plot as $\mathcal{A}$ and $\mathcal{B}$, to implement a phase gate. Path $\mathcal{A}$ modulates the amplitude of the base drive tone, while path $\mathcal{B}$ modulates the base drive tone plus a second drive tone to remain in a region with less decoherence. To induce the phase gate, the system starts in point $\mathcal{O}$, which is a double sweet spot and has the maximal dephasing time with respect to $A_m$. During the gate, the drive parameters are modulated using a Gaussian pulse, following either the path $\mathcal{A}$ or $\mathcal{B}$. At the end of the pulse, the drive returns to point $\mathcal{O}$. This modulation of the driving parameters along paths $\mathcal{A}$ (red lines) and $\mathcal{B}$ (blue lines) is shown in Fig.~\ref{fig:PhaseGate}(b), with the dashed and solid lines representing $A_m(t)$ and $A_n(t)$, respectively. The modulation of the drive parameters for path $\mathcal{B}$ are obtained by maximizing $T_2$ during the pulse. In this scheme, the drive frequency is tuned to remain at the dephasing peak throughout the pulse (thus shifting slightly throughout the pulse). Note that this tuning can also be done by keeping the drive frequency fixed, but slowly varying flux $\phi_\mathrm{dc}$ of the fluxonium instead.

The angle of rotation around the $z$-axis of the Bloch sphere is related to the change in the quasi-energy splitting,
\begin{equation}
    \theta = \int_{0}^{\tau_{g}} \Delta\epsilon_\mathrm{gate}(t)\text{d}t,
\end{equation}
with $\Delta\epsilon_\mathrm{gate}(t) = \Delta\epsilon(t) - \Delta\epsilon(0)$. The change in the quasi-energy splitting is plotted in Fig.~\ref{fig:PhaseGate}(d). The modulation amplitudes of $A_m$ and $A_n$ were chosen such that the angle of rotation is $\pi$. During the gate pulse, $T_2^\mathrm{peak}$ is lowered along path $\mathcal{A}$, while it is increased along path $\mathcal{B}$, as shown in Fig.~\ref{fig:PhaseGate}(e). This is due to the choice of the location of $\mathcal{O}$ at a double sweet spot. As a result, modulating only the base drive tone inevitably decreases $T_2^\mathrm{peak}$, while modulating the second drive tone allows the system to remain at the double sweet spot during the gate pulse. Simulations of these two gates were performed and show the average gate error to be decreased by a factor two when using the second drive. An example of such a simulation is shown in Fig.~\ref{fig:PhaseGate}(c), where the qubit state was initialized in $\ket{-}=\ket{v_{+,0}(t)}-\ket{v_{-,0}(t)}$ with $\ket{v_{\pm,0}(t)}$ the computational basis states at drive point $\mathcal{O}$. The projection on $\ket{+}=\ket{v_{+,0}(t)}+\ket{v_{-,0}(t)}$ is plotted during the gate pulse, which reveals different paths for the different gate paths. This effect is due to small variations in the Floquet modes throughout the gate pulse.

\section{Conclusion}

\label{Conclusion}

We present an analytical and numerical analysis of coherence times in fluxonium qubits driven by two commensurable tones, with a drive form given by $\phi_{ac}(t)=\phi_m\cos(m\omega_\mathrm{d} t)+\phi_n\cos(n\omega_\mathrm{d} t + \varphi)$. We show that, using this two-tone driving scheme, the coherence times can be further optimized by operating the qubit at dephasing triple sweet spots, which exhibit higher dephasing times and wider dephasing peaks due to a lowered sensitivity with respect to noise on the static magnetic flux and the amplitude of the two drive tones.
Analyzing different two-tone drive forms, we find that adding a higher harmonic to a base drive tone is ideal to increase dephasing times beyond the maximal dephasing time that can be achieved with a single base tone.
Following the single-qubit phase-gate protocol of Ref.~\cite{huang2021engineering}, we demonstrate that a two-tone pulse can implement the phase gate with a higher fidelity as compared to a single-tone pulse, due to higher dephasing times throughout the pulse.

\appendix

\section{Derivation of the two-level Hamiltonian}
\label{app:qubithamiltonian}
The fluxonium qubit consisting of a Josephson junction with tunneling energy $E_\mathrm{J}$, inductance with inductive energy $E_\mathrm{L}$ and capacitance with charging energy $E_\mathrm{C}$ can be described by the quantized circuit Hamiltonian
\begin{equation}
    \hat{H}_{\mathrm{c}} = 4E_\mathrm{C}\hat{n}^2 + \frac{1}{2}E_\mathrm{L}(\hat{\phi} - \phi_\mathrm{ext})^2 - E_\mathrm{J}\cos(\frac{2\pi}{\phi_0}\hat{\phi}),
    \label{eq:cqed}
\end{equation}
with the capacitor charge operator $\hat{n}$ and $\hat{\phi}$ the phase operator. $\phi_\mathrm{ext}$ is the externally applied magnetic flux and $\phi_0$ the flux quantum. Since the qubit is biased near the anticrossing at $\phi_\mathrm{ext}=\phi_0/2$, a two-level approximation can be made using the eigenstates of $\ket{0}$ and $\ket{1}$ of $\hat{H}_\mathrm{c}$ when $\phi_{\mathrm{ext}} = \phi_0/2$,
\begin{equation}
    \hat{H} = \begin{bmatrix}
        \bra{0}\hat{H}_{\mathrm{c}}\ket{0}&\bra{0}\hat{H}_\mathrm{c}\ket{1}\\\bra{1}\hat{H}_{\mathrm{c}}\ket{0}&\bra{1}\hat{H}_{\mathrm{c}}\ket{1}
    \end{bmatrix} = 
    \begin{bmatrix}
        E_0&-B/2\\-B/2&E_1
    \end{bmatrix}
\end{equation}
with $B = 2\frac{\phi_\mathrm{ext} - \phi_0/2}{L}\bra{0}\phi\ket{1}$. This can be rewritten in terms of the Pauli matrices,
\begin{equation}
    \hat{H} = \frac{\Delta}{2}\hat{\sigma}_z - \frac{B}{2}\hat{\sigma}_x,
\end{equation}
with $\Delta = E_1-E_0$. This Hamiltonian is then rotated using the unitary $\hat{U} = e^{-i\frac{\pi}{4}\hat{\sigma}_\mathrm{y}}$ to become
\begin{equation}
    \hat{H} = \frac{B}{2}\hat{\sigma}_z + \frac{\Delta}{2}\hat{\sigma}_x.
\end{equation}
In the case of a two-tone drive, the externally applied magnetic flux is of the form $\phi_\mathrm{ext}(t) = \phi_\mathrm{ext} + \phi_{m}\cos(m\omega_\mathrm{d} t) + \phi_{n}\cos(n\omega_\mathrm{d} t)$ and the truncated Hamiltonian reads
\begin{equation}
    \hat{H}(t) = \frac{\Delta}{2}\hat{\sigma}_x + \frac{B}{2}\hat{\sigma}_z + A_m\cos(m\omega_\mathrm{d} t)\hat{\sigma}_z + A_n\cos(n\omega_\mathrm{d} t)\hat{\sigma_z},
\end{equation}
with $A_{m,n} = E_\mathrm{L}\bra{0}\phi\ket{1}\phi_{m,n}$ which is the Hamiltonian used in the text.

\section{Gap size approximation in the strong-drive limit}
\label{App:GapSize}
 In the strong-drive limit, $\frac{\Delta}{2}\hat{\sigma}_x$ is treated as a perturbation to the unperturbed Hamiltonian \cite{huang2021engineering, PhysRevA.79.032301}
 \begin{equation}
     \hat{H}_0(t) = \bigg( \frac{B}{2}+A_m\cos(m\omega_\mathrm{d}t)+A_n\cos(n\omega_\mathrm{d}t+\varphi) \bigg)\hat{\sigma}_z.
 \end{equation}
Plugging $\hat{H}_0(t)$ into the Floquet equation yields the Floquet modes
\begin{multline}
    \ket{v_{\pm, s}^{(0)}(t)} = \exp\bigg(\mp i\frac{A_m}{m\omega_\mathrm{d}}\sin(m\omega_\mathrm{d}t) \\ \mp i\frac{A_n}{n\omega_\mathrm{d}}\sin(n\omega_\mathrm{d}t+\varphi) + is\hbar\omega_\mathrm{d}t\bigg)\ket{\pm_z},
\end{multline}
where $\ket{\pm_z}$ are the eigenstates of the $\sigma_z$-operator. The quasi-energy levels associated with these modes are $\epsilon_{\pm, s} = \pm B/2 +s\omega_\mathrm{d}$, which exhibit crossings when $\omega_\mathrm{d} = B/s$. When the perturbation in turned on, this degeneracy is lifted and the crossings become avoided crossings with a gap size of $\Delta_s$. Using first order degenerate perturbation theory,
\begin{equation}
    \Delta_s \approx \Delta\abs{\frac{\omega_\mathrm{d}}{2\pi}\int_0^{2\pi/\omega_\mathrm{d}}dt \bra{v_{+, 0}^{(0)}(t)}\hat{\sigma_x}\ket{v_{-,s}^{(0)}(t)}},
\end{equation}
which, using the Jacobi-Anger expansion, can be rewritten as
\begin{equation}
    \Delta_s \approx \Delta\abs{\sum_{(k, k')}J_k\bigg(\frac{2A_m}{m\omega_\mathrm{d}}\bigg)J_{k'}\bigg( \frac{2A_n}{n\omega_\mathrm{d}} \bigg)e^{ik'\varphi}}.
\end{equation}
The summation runs over all $(k,k')$ which are integer solutions to the equation $mk+nk'=s$, due to $\frac{\omega_\mathrm{d}}{2\pi}\int_0^{2\pi/\omega_\mathrm{d}}dt\text{ }e^{i(mk+nk'-s)\omega_\mathrm{d}t} = \delta_{mk+nk',s}$.

\section{Noise model}
\label{app:noise}

The noise considered in this work consists of several parts \cite{PhysRevApplied.20.034016}. First, there is noise on the static magnetic flux $\phi_\mathrm{ext}$ \cite{PhysRevB.72.134519}, with a power spectral density given by
\begin{equation}
    S_\text{dc}(\omega)=2\pi\frac{A_\text{dc}^2}{\abs{\omega}},
\end{equation}
with $A_\text{dc} = 2\pi\delta_\mathrm{f}E_\mathrm{L}|\bra{0}\hat{\phi}\ket{1}|$, with flux noise amplitude $\delta_\mathrm{f} = 1.8\times10^{-6}$. This noise couples to the phase operator, $\hat{\phi}$, in the circuit Hamiltonian in Eq.~\eqref{eq:cqed}. Dielectric loss is also taken into account \cite{PhysRevX.9.041041}, which also couples to $\hat{\phi}$ and is described by the power spectral density
\begin{equation}
    S_\text{dl}(\omega)=\frac{1}{2}A_\text{dl}\abs{\coth(\frac{\omega}{2k_\text{B}T})+1}\bigg(\frac{\omega}{2\pi}\bigg)^2,
\end{equation}
with $A_\text{dl}=\pi^2\tan\delta_\mathrm{C}|\bra{0}\hat{\phi}\ket{1}|^2/E_\mathrm{C}$ and loss tangent $\tan\delta_\mathrm{C} = 1.1\times10^{-6}$. In the two-level Hamiltonian of Eq.~\eqref{eq:system} both of these noise sources couple to the $\hat{\sigma}_z$-operator \cite{PhysRevX.9.041041}. The noise parameters chosen here are typical values found in experiments \cite{PhysRevX.9.041041, PhysRevX.11.011010}. The driving also adds a source of noise to the system. While the drive frequency can be kept very stable \cite{PhysRevApplied.14.054033}, the amplitudes are expected to be noisy.  Following Ref.~\cite{huang2021engineering}, we assume the following 1/$f$ noise spectra for the drive amplitudes \cite{PhysRevApplied.12.054015, PhysRevA.101.012302},
\begin{equation}
    S_\text{ac}(\omega) = 2\pi\frac{A_\text{ac}^2}{\abs{\omega}},
\end{equation}
with $A_\text{ac}=0.9 A_\text{dc}$ \cite{PhysRevApplied.14.054033}. This noise couples to the operators $\cos(m\omega_\mathrm{d}t)\hat{\sigma}_z$ and $\cos(n\omega_\mathrm{d}t+\varphi)\hat{\sigma}_z$ with the corresponding frequencies $m\omega_\mathrm{d}$ and $n\omega_\mathrm{d}$.

\section{Decoherence rates}
\label{dec_rates}
The decoherence rates are derived using Eq.~\eqref{Hint} in a Bloch-Redfield equation for the qubit-bath system, using the same procedure as in Refs.~\cite{huang2021engineering, PhysRevApplied.14.054033}. The resulting rates are
\begin{align}
\begin{split}
    \gamma_{\pm} = \sum_{k\in\mathbb{Z}}|g_{k}^\pm|^2\bigg(S_\text{dc}(k\omega_\mathrm{d}\pm\epsilon_{01})+S_\text{dl}(k\omega_\mathrm{d}\pm\epsilon_{01})\bigg)\\+\sum_{k\in\mathbb{Z}}\abs{g_{k-m}^\pm+g_{k+m}^\pm}^2S_\text{ac,m}(k\omega_\mathrm{d}\pm\epsilon_{01})\\+\sum_{k\in\mathbb{Z}}\abs{g_{k-n}^\pm+g_{k+n}^\pm}^2S_\text{ac,n}(k\omega_\mathrm{d}\pm\epsilon_{01}),
\end{split}
\end{align}
for relaxation, and
\begin{align}
\begin{split}
    \gamma_{\phi} = \sum_{k\in\mathbb{Z}}2|g_{k}^\phi|^2\bigg(S_\text{dc}(k\omega_\mathrm{d})+S_\text{dl}(k\omega_\mathrm{d})\bigg)\\+\sum_{k\in\mathbb{Z}}\frac{1}{2}\abs{g_{k-m}^\phi+g_{k+m}^\phi}^2S_\text{ac,m}(k\omega_\mathrm{d}) \\+\sum_{k\in\mathbb{Z}}\frac{1}{2}\abs{g_{k-n}^\phi+g_{k+n}^\phi}^2S_\text{ac,n}(k\omega_\mathrm{d}),
\end{split}
\end{align}
for dephasing, with $g^\pm_k$ and $g^\phi_k$ as defined in Eq.~\eqref{gterms1} and Eq.~\eqref{gterms2}, respectively.

\section{Choice of $m$ and $n$}
\label{mnchoice}
\begin{figure}
    \centering
    \includegraphics[width=\linewidth]{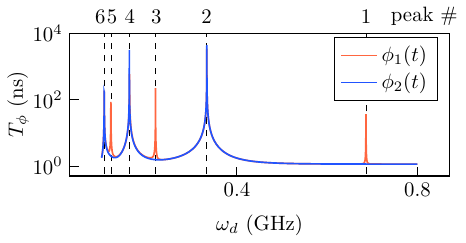}
    \caption{Peaks in the dephasing time for a range of drive frequencies $\omega_\mathrm{d}$, with $\phi_1(t) = 0.22\cos(2\omega_\mathrm{d}t) - 0.01\cos(3\omega_\mathrm{d}t)$ and $\phi_2(t) = 0.22\cos(2\omega_\mathrm{d}t)$ the drive forms of the magnetic flux. When driving with $\phi_2(t)$, only the even peak numbers are present, as the drive frequency is $2\omega_\mathrm{d}$. When turning on the second drive at frequency $3\omega_\mathrm{d}$, the total drive frequency becomes $\omega_\mathrm{d}$, and the peaks with odd peak number start to appear.}
    \label{fig:peak_numbers}
\end{figure}

Since we require $\left.\frac{\partial\Delta_s}{\partial A_n}\right|_{A_n=0}\neq 0$ for the derivative condition, the summation in Eq.~\eqref{eq:delta_s} has to contain $k'=\pm1$, as explained in Sec.~\ref{section 4}. The summation runs over $k$ and $k'$ which satisfy
\begin{equation}
    mk+nk'=s,
\end{equation}
with $m, n$ the chosen drive parameters, and $s\in\mathbb{N}$ the number of the relevant anticrossing. When $k'=\pm1$, this becomes
\begin{equation}
    mk\pm n=s.
    \label{eq:condition}
\end{equation}
There are two distinct situations concerning the drive parameters. The first is when the drive is made up of a base frequency and a harmonic, i.e. $m=1$ and $n\neq1$. In this case, we get
\begin{equation}
    k\pm n = s,
\end{equation}
which has solutions of the form $(k,k') = (s\mp n, \pm1)$. It is clear that the derivative condition is met for every peak number $s$, since there is an integer solution of $k$ for all possible values of $s$. The second case is when neither $m$ nor $n$ is equal to 1, then Eq.~\eqref{eq:condition} cannot be reduced further and has solutions of the form $(k,k') = (\frac{s\mp n}{m}, \pm1)$ The derivative condition can thus only be met for peak numbers $s$ for which $s\mp n$ is evenly divisible by $m$. Furthermore, the peaks for which this is the case do not exist at $A_n=0$. To see this, remember that for a single frequency drive of frequency $m\omega_\mathrm{d}$, there are only peaks with peak number $ms'$, with $s'\in\mathbb{N}$. This is illustrated in Fig.~\ref{fig:peak_numbers}. Thus, at peaks where $A_n=0$, Eq.~\eqref{eq:condition} becomes
\begin{equation}
    mk\pm n=ms',
\end{equation}
which has no solutions in $k$ and $s'$, since $k=s'\mp n/m$, and $\text{gcd}(m,n)=1$.

\section{$A_m$ and $A_n$ sweet spots}
\label{noisy_drive}
The sensitivity of the driven qubit with respect to noise on the drive amplitudes can be approximated by applying a small perturbation to the Hamiltonian, $\Bar{H}\rightarrow\Bar{H}+\delta A_i\Bar{\sigma}_{z,i}$, from which a first-order correction to the quasi-energy difference $\Delta\epsilon$ can be derived \cite{PhysRevApplied.14.054033, huang2021engineering},
\begin{equation}
    \delta\Delta\epsilon = \delta A_i (\bra{v_1}\Bar{\sigma}_{z,i}\ket{v_1} - \bra{v_0}\Bar{\sigma}_{z,i}\ket{v_0}).
\end{equation}
When evaluated, this yields
\begin{equation}
    \frac{\delta\Delta\epsilon}{\delta A_i} = g_{i,\phi} + g_{-i, \phi},
\end{equation}
with $g_{i,\phi}$ the $i^\text{th}$ dephasing coupling matrix element. Finding a triple sweet-spot thus amounts to finding a point in the driving-parameter space where $g_{0, \phi} = g_{m, \phi} = g_{-m, \phi} = g_{n, \phi} = g_{-n, \phi}=0$, which is the location where the partial derivatives of $\Delta\epsilon$ with respect to $B$, $A_m$ and $A_n$ equal zero.

\bibliography{refs}

\end{document}